\documentclass[sigconf]{acmart}

\usepackage{subcaption}
\usepackage{cleveref}
\usepackage{enumitem}
\usepackage{xspace}
\usepackage{float}
\usepackage{multirow}
\usepackage{balance}

\AtBeginDocument{%
  }

\copyrightyear{2026}
\acmYear{2026}
\setcopyright{cc}
\setcctype{by}
\acmConference[KDD 2026]{Proceedings of the 32nd ACM SIGKDD Conference on Knowledge Discovery and Data Mining V.2}{August 9--13, 2026}{Jeju Island, Republic of Korea}
\acmBooktitle{Proceedings of the 32nd ACM SIGKDD Conference on Knowledge Discovery and Data Mining V.2 (KDD 2026), August 9--13, 2026, Jeju Island, Republic of Korea}
\acmISBN{979-8-4007-2259-2/2026/08}
\acmDOI{10.1145/3770855.3818462}
\newcommand{\pinrec}{PinRec\xspace}

\newcommand{\aref}[1]{\hyperref[#1]{Appendix~\ref*{#1}}}

\begin{document}
\title{PinRec: Unified Generative Retrieval for Pinterest Recommender Systems}


  \author{Edoardo Botta}
  \orcid{0009-0003-6246-8477}
  \authornote{Work done at Pinterest.}
  \affiliation{%
    \institution{Pinterest}
    \city{Palo Alto}
    \state{CA}
    \country{USA}
  }
  \email{ebotta@pinterest.com}

  \author{Jaewon Yang}
  \orcid{0009-0001-2224-7915}
  \authornote{Corresponding author.}
  \affiliation{%
    \institution{Pinterest}
    \city{Palo Alto}
    \state{CA}
    \country{USA}
  }
  \email{jaewonyang@pinterest.com}

  \author{Yi-Ping Hsu}
  \orcid{0009-0002-3764-3643}
  \affiliation{%
    \institution{Pinterest}
    \city{Palo Alto}
    \state{CA}
    \country{USA}
  }
  \email{yhsu@pinterest.com}

  \author{Laksh Bhasin}
  \orcid{0009-0007-8734-4330}
  \affiliation{%
    \institution{Pinterest}
    \city{Palo Alto}
    \state{CA}
    \country{USA}
  }
  \email{lbhasin@pinterest.com}

  \author{Yilin Chen}
  \orcid{0000-0001-9850-7157}
  \affiliation{%
    \institution{Pinterest}
    \city{New York}
    \state{NY}
    \country{USA}
  }
  \email{yilinchen@pinterest.com}

  \author{Prabhat Agarwal}
  \orcid{0000-0002-3826-0858}
  \authornotemark[1]
  \affiliation{%
    \institution{Pinterest}
    \city{Palo Alto}
    \state{CA}
    \country{USA}
  }
  \email{pagarwal@pinterest.com}

  \author{Anirudhan Badrinath}
  \orcid{0000-0003-4572-4566}
  \authornotemark[1]
  \affiliation{%
    \institution{Pinterest}
    \city{Palo Alto}
    \state{CA}
    \country{USA}
  }
  \email{abadrinath@pinterest.com}

  \author{Jiajing Xu}
  \orcid{0000-0002-4761-5171}
  \affiliation{%
    \institution{Pinterest}
    \city{Palo Alto}
    \state{CA}
    \country{USA}
  }
  \email{jiajing@pinterest.com}

  \author{Charles Rosenberg}
  \orcid{0009-0003-9664-8644}
  \affiliation{%
    \institution{Pinterest}
    \city{Palo Alto}
    \state{CA}
    \country{USA}
  }
  \email{crosenberg@pinterest.com}

\renewcommand{\shortauthors}{Edoardo Botta et al.}

\begin{abstract}
Generative retrieval methods employ sequential modeling techniques, like transformers, to generate candidate items for recommender systems. These methods have demonstrated promising results in academic benchmarks, surpassing traditional retrieval models such as two-tower architectures. However, a key limitation is that current approaches require a separate model for each product surface, as building a unified model that accommodates the different business needs of various surfaces has proven challenging. Furthermore, existing methods often fail to capture the evolution of user interests over a sequence, focusing instead on only predicting the next item.

This paper introduces \pinrec, a novel unified generative retrieval model for all of Pinterest’s recommendation surfaces, including home feed, search, and related pins. \pinrec is pretrained on user activity sequences aggregated across surfaces, then fine-tuned for each surface using that surface’s impression data. This pretraining–fine-tuning approach enables a single unified model while still adapting to the needs of individual surfaces. To better align recommendations with surface-specific business goals, \pinrec incorporates a novel outcome-conditioned generation mechanism that targets different outcomes for each surface, which further enhances the impact of fine-tuning. Our experiments show that \pinrec balances performance, diversity, and efficiency, delivering significant gains such as +4\% increase in search saves. To our knowledge, this paper presents the first rigorous study of a unified generative retrieval model built and deployed at Pinterest scale, marking a significant milestone in the field.

\end{abstract}




\maketitle

\begin{figure}[ht]
    \centering
    \includegraphics[width=1.06\linewidth]{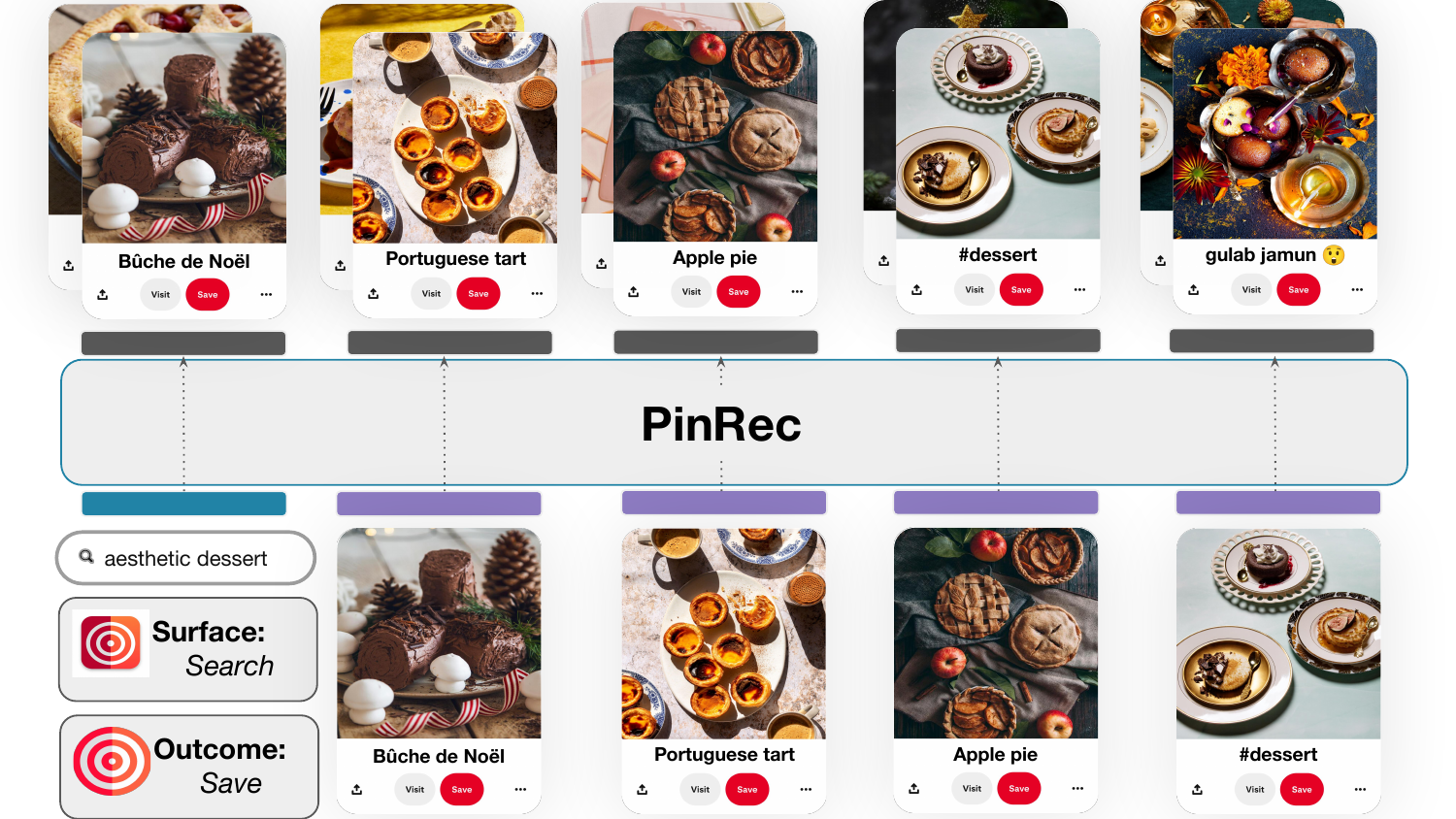}
    \caption{Illustration of \pinrec, a unified generative  retrieval model for heterogeneous user journeys on Pinterest. Sequences of user searches and activities are used to recommend Pins (top). \pinrec powers retrieval in multiple surfaces optimizing surfaces' unique business metrics like Saves for Home Feed and Product Click for Search}
    \vspace{-1mm}
    \label{fig:cover_graphic}
\end{figure}

\section{Introduction}
\label{sec:intro}

%
Pinterest is a visual discovery platform where over 600 million monthly active users search for, save, and shop for ideas, represented by Pins~\cite{pins2025}. Given the massive volume of Pins (for instance, users save over 1.5 billion Pins per week~\cite{pins2}), it is essential to provide personalized recommendations in Pinterest feeds and search results. To achieve this, recommender systems often leverage a two-stage approach: retrieval and ranking. Retrieval models select thousands of potentially relevant candidate Pins, then ranking models prioritize the top Pins to display to the user. Since ranking models depend on the retrieval results, it is important for retrieval models to identify diverse, fresh, and relevant Pins.

Traditionally, retrieval methods employ ``two-tower'' models. The query tower computes query embeddings, while the item tower computes Pin embeddings. These models retrieve Pins with the highest similarity scores given a query embedding. As an alternative to two-tower models, recent work has proposed \emph{Generative Retrieval (GR)}. GR models leverage a sequence of tokens using generative models, such as Transformers~\citep{vaswani2017attention}, and then use those tokens to retrieve Pins. GR has shown promising results in academic settings given its powerful ability to understand user activity history and generate candidates that existing two-tower models may not be able to retrieve~\cite{KangM18, Recformer23, UnifyingGenRetrieval24}.

In this paper, we aim to develop a \emph{unified} GR model that can serve multiple Pinterest recommendation surfaces (e.g., search, home feed, related Pins) by being fine-tuned on the target surface. Because GR models operate over a single token sequence, they can combine heterogeneous inputs (such as query text and user history) into one representation, enabling parameter sharing and cross-surface transfer. Moreover, the training signal is naturally cross-surface: interaction logs from different surfaces can be expressed in a common next-item (or next-token) prediction framework, allowing the model to learn from a unified mixture of surface-specific trajectories. In contrast, two-tower systems often require surface-specific query towers, objectives, and/or retrieval indices, which increases complexity and can limit generalization across surfaces.

Despite these advantages, translating a unified GR approach into a production retriever that serves multiple Pinterest surfaces introduces several practical challenges. First, it must efficiently optimize \emph{surface-specific objectives}, since each surface defines success differently: search emphasizes outbound clicks that reflect offsite conversion intent, related Pins emphasizes session depth and continued exploration, and home feed emphasizes saves and discovery (clicks and downstream exploration). This is further complicated when key metrics are sparse; for example, outbound product clicks in search are relatively rare events. Second, user behavior and demographics differ by surface: upstream surfaces at the start of a session (e.g., the home feed) have a higher share of less-active users, whereas downstream surfaces (e.g., related Pins) skew toward more active users who take stronger actions. Third, user intent can evolve even within a single session, and a generative retriever must adapt its generated sequence accordingly. Fourth, due to sequential decoding, GR models typically incur higher computational cost and latency than two-tower models, which only require embedding computation and nearest-neighbor search.

\begin{figure}[!t]
    \centering
    \includegraphics[width=1.05\linewidth, trim=0 20 0 0, clip]{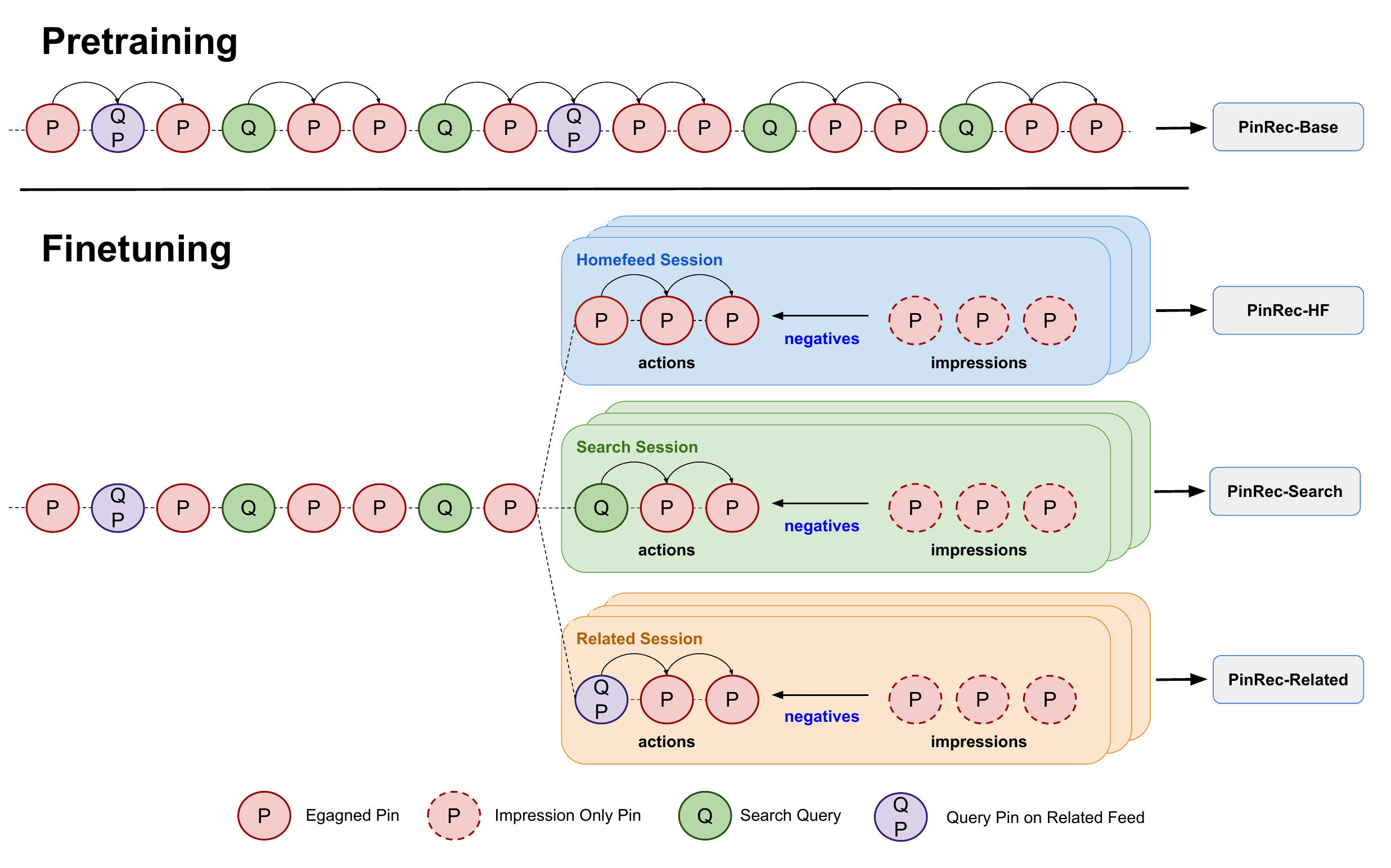}
    \caption{\pinrec is pretrained with the user activity sequences consisting of items (Pins) and search queries. \pinrec is then fine-tuned for each surface specific data that contains user actions and impressed items (Pins). }
    \vspace{-0mm}
    \label{fig:training}
\end{figure}

To our knowledge, no prior work addresses \emph{all} of these requirements in a single unified GR model, namely (i) \emph{surface-specific} success metrics, (ii) heterogeneity in cross-surface data distributions, (iii) in-session user intent shifts, and (iv) the latency/cost constraints. Recent industrial GR systems such as TIGER and HSTU~\cite{HSTU24, Tiger23} are designed for a single next-token/next-item prediction objective, and in practice are typically trained against a unified engagement target (e.g., any action) rather than explicitly aligning the learning signal to surface-specific goals (e.g., outbound clicks for search, saves for home feed). As a result, fine-tuning can yield limited gains when the surface ``true north'' metrics differ (see Sec.~\ref{sec:oc_fine_tuning}). Separately, foundational sequence models like RecFormer and PLUM can be fine-tuned for individual applications~\cite{UniS4Rec22, Recformer23, Plum25} and can better match surface-specific user/item distributions, but they are often prohibitively expensive at Pinterest scale due to their reliance on LLM backbones; moreover, they are trained for a single generic engagement target, leaving the objective-misalignment issue unchanged. Finally, existing GR approaches do not support multi-step generation at retrieval time; because autoregressive decoding is costly, they commonly generate only the next-item candidates, limiting their ability to produce a longer candidate sequence that anticipates or tracks within-session shifts in intent.


\begin{figure}[t]
    \centering
    \includegraphics[width=1.1\linewidth, trim=40 20 0 0, clip]{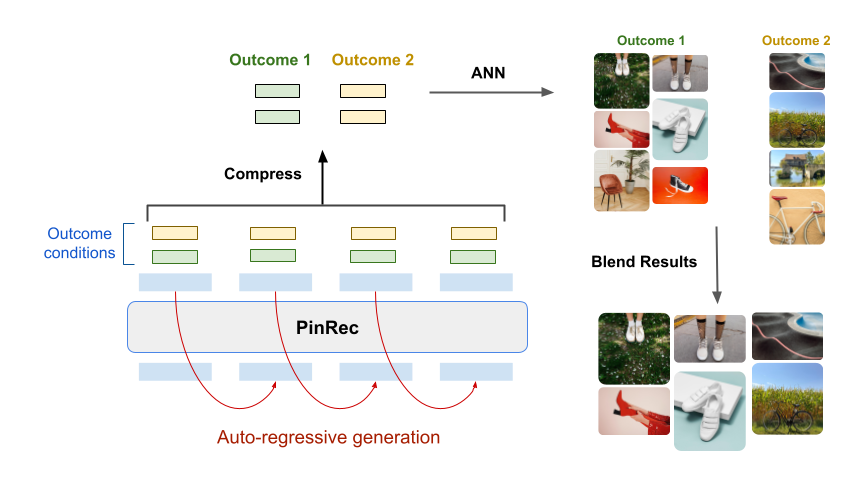}
    \vspace{-2mm} 
    \caption{Generation phase of \pinrec}
    \vspace{-1mm}
    \label{fig:model}
\end{figure}

In this paper, we present \pinrec, a foundational GR model for Pinterest recommendation surfaces. We pretrain \pinrec on user activity data aggregated across surfaces~\cite{PinFM25}, and then fine-tune it on surface-specific data to align the model with each surface’s business goals (Fig. ~\ref{fig:training}). Unlike prior GR methods, \pinrec addresses four aforementioned challenges: (1) optimizing surface-specific objectives, (2) adapting to each surface’s data distribution, (3) capturing evolving user intent within a session via sequence generation, and (4) doing so with only a marginal increase in serving cost. We summarize our solutions to each challenge below.

\begin{itemize}[align=parleft,left=0pt..1em]
    \item \textbf{Outcome-Conditioned Generative Retrieval:} To optimize surface-specific objectives, \pinrec employs \textit{outcome-conditioned generation}, enabling the optimization of multiple business metrics. For each outcome (e.g., user saves or clicks), \pinrec generates a \emph{sequence} of Pin tokens conditioned on that outcome and retrieves candidate Pins; the generated sequences can then be combined to balance objectives according to desired trade-offs (Fig.~\ref{fig:model}).  Our offline and online evaluations show that outcome conditioning enables \pinrec to adapt to the distinct business needs of each surface.

    \item \textbf{Fine-tuning for each surface:} To adapt for each surface, we employ a pretraining-fine-tuning framework. We first pretrain the base \pinrec model on user activity sequences aggregated across all Pinterest surfaces. To tailor the model to each surface’s user population and behavior patterns, we then fine-tune \pinrec on surface-specific impression logs. Using this approach, we deployed \pinrec in production across three surfaces. To our knowledge, this is the first paper to apply a pretraining-fine-tuning paradigm for generative retrieval across multiple recommendation surfaces at web scale.

    \item \textbf{Efficient Serving with Auto-regressive generation:} To capture evolving user interest, \pinrec generates item sequences auto-regressively, with up to +71\% recall gains from multi-step generation (Table~\ref{table:ar_ablation}). To mitigate the cost of generation, we describe optimizations that enable low-latency decoding, detail the serving infrastructure, and show how we scale the system horizontally to serve hundreds of millions of users. To our knowledge, this is the first study to disclose infrastructure optimizations for an industry-scale generative retrieval system.
\end{itemize}

We will make the model code publicly available upon publication.


\section{Related Work}

\textbf{Sequential Recommendation Systems:} Sequential recommendation systems operate on a sequences of items with which a user has interacted~\cite{KangM18}. Earlier work in academia has used recurrent neural networks (RNN)~\cite{GRU4REC16}, but later work has often adopted transformers~\cite{KangM18, BERT4REC19}. Recent studies have harnessed the extensive knowledge of pre-trained language models~\cite{UniS4Rec22, Recformer23, Plum25}.

In industrial applications, a common approach is to apply sequential encoders to learn user representations~\cite{Pinnerformer22}. The user representation can then be used in traditional recommender systems (e.g., embedding-based retrieval or ranking models)~\cite{Spotify20, rethink, Transact23, TWINV224}.
A different line of work in sequential recommendation is \emph{generative recommendation}, where the model directly generates the items to recommend. These methods fall into two paradigms based on output representation. \emph{Token-ID approaches} (e.g., TIGER~\citep{Tiger23}, OneRec~\citep{Onerec25}) decode sequences of discrete semantic ID tokens~\cite{RQVAE22, SemanticIDGeneral24} learned via vector quantization. While effective on academic benchmarks, they face challenges at industrial scale: hallucination of invalid ID sequences, the hourglass phenomenon where many items collapse to identical IDs~\citep{Hourglass24, MultiBehaviorSASRec24}, and the need for separate decoding infrastructure. \emph{Embedding-based approaches} (e.g., LIGER~\citep{UnifyingGenRetrieval24}, HSTU~\citep{HSTU24}) instead generate continuous representations and retrieve items via approximate nearest neighbor (ANN) search, avoiding hallucination by construction and integrating with existing serving infrastructure. \pinrec follows the embedding-based paradigm, where multi-surface pretraining unifies heterogeneous interactions into a single sequence and continuous output enables retrieval over any item corpus using shared ANN indices.

Very recently, \citet{Onerec25, Onepiece25} proposed multi-stage training frameworks that unify retrieval and ranking within a single surface. This represents \emph{vertical unification}, integrating different stages of the recommendation funnel for one surface. In contrast, we pursue \emph{horizontal unification}: we fine-tune a single retrieval model to serve multiple surfaces. Both directions are important, but we expect horizontal unification to be easier to translate into product impact, since the unified model can be integrated into existing multi-stage funnels as a drop-in replacement at the retrieval stage.


\textbf{Multi-Objective Generation:} We discuss generative models that consider multiple objectives. While RL algorithms have shown this capability in industry \citep{SelfsupervisedRL20}, studies in language modeling suggest RL often needs extensive tuning for optimal generation \citep{PPO22}. Recent research has introduced more streamlined and scalable techniques based on supervised learning, such as goal-conditioned methods~\cite{IterativeSupervised21} and prompting strategies~\cite{PromptingRL22}. \pinrec builds on these approaches to enhance scalability.

\section{Problem Setup}
\label{sec:setup}

We formulate the generative sequential recommendation task as follows. Consider a system with users $\mathcal{U}$, items $\mathcal{I}$, actions $\mathcal{A}$, and surfaces $\mathcal{S}$. Our objective is to generate an ordered list of $N$ item recommendations $R(u, t) \subset \mathcal{I}$ for user $u \in \mathcal{U}$ at timestep $t$.

\textbf{Item Types and Actions.} Each item $i \in \mathcal{I}$ belongs to an item type $T(i) \in \mathcal{T}$, where $\mathcal{T} = \{T(i) \mid i \in \mathcal{I}\}$ represents the set of all item types in the system (e.g., Pins, boards, search queries). Users can perform various actions $\mathcal{A} = \{\text{click}, \text{save}, \text{share}, \text{comment}, \ldots\}$ on these items across different surfaces. Each interaction occurs on a specific surface $s \in \mathcal{S}$, where surfaces represent different recommendation products, such as homefeed, search results, board pages, and related pin feed.

\textbf{User History.} For each user $u$, we maintain their interaction history $H(u, t_{\text{max}}) = \langle (i_1, a_1, s_1, t_1), (i_2, a_2, s_2, t_2), \ldots, (i_m, a_m, s_m, t_m) \rangle$, where each tuple $(i_j, a_j, s_j, t_j)$ represents an interaction with item $i_j \in \mathcal{I}$ via action $a_j \in \mathcal{A}$ on surface $s_j \in \mathcal{S}$ at timestep $t_j \leq t_{\text{max}}$. The sequence is chronologically ordered, $t_1 \leq t_2 \leq \ldots \leq t_m$.

\textbf{Surface-specific Feed View.} When user $u$ visits surface $s$ at time $t$, we observe a feed-view event
$F(u,s,t)=\langle (i_1,a_1, t_1), \ldots, (i_m,a_m, t_m)\rangle,$
where each tuple $(i_j,a_j,t_j)$ indicates that item $i_j\in\mathcal I$ was shown to the user at $j$-th position, and the user’s action on that impression is encoded by $a_j \in \mathcal{A}\cup\{\text{imp}\},$
with $\text{imp}$ denoting an \emph{impression with no explicit action taken} (i.e., the item was shown but not clicked/saved/etc.). If an explicit engagement occurs (e.g., click, save), then $a_j\in\mathcal A$. The timestamp $t_j$ is the time of action when $a_j\in\mathcal A$; otherwise (no action), we set $t_j = t$.

\textbf{Sequential Recommendation Model.} We propose a parameterized sequence model $f_\theta: \mathcal{H} \times \mathcal{T} \times \mathcal{S} \rightarrow \mathbb{R}^{|\mathcal{I}|}$ that maps a user's interaction history, item types, and surface context to item scores, where $\mathcal{H}$ denotes the space of all possible interaction histories. The model parameters $\theta$ are optimized over the collection of user interaction sequences $\{H(u, t_{\text{max}}) \mid u \in \mathcal{U}\}$ to maximize the likelihood of observed user interactions while accounting for the heterogeneous nature of actions and surfaces. This enables the model to learn nuanced behavioral patterns that depend on both what users interact with and how and where they interact with it.




\section{\pinrec}
\label{sec:pinrec}

In this section, we describe the model architecture and serving infrastructure for \pinrec. We start with an overview of the model, including outcome-conditioned generation. Next, we explain how items in the sequence are encoded and processed. Then we talk about how we perform pretraining on a unified data and how we fine-tune for a surface-specific data. Finally, we describe inference mode operation and provide details about our real-time serving architecture, which is powered by Nvidia Triton.

\subsection{Model Architecture}

We use a transformer decoder architecture as the backbone of our \pinrec\ model. The user interaction history is first featurized and fed into a stack of transformer layers along with learned position embeddings to capture temporal ordering. At each layer, we apply causal attention masking to ensure that tokens at time $t$ can only attend to tokens at positions $t' \leq t$, preventing information leakage from future interactions. The stack of $L$ transformer layers outputs a sequence of contextualized hidden states: $ \mathbf{h}_{u,t} = \text{TransformerStack}^{(L)}(\mathbf{x}_{u,1:t}) $.

where $\mathbf{x}_{u,1:t}$ represents the featurized user interaction sequence up to timestep $t$, and $\mathbf{h}_{u,t}$ is the resulting hidden state at position $t$ from the final layer. These hidden states are then processed through an output head $O$ to generate the next item predictions.

\textbf{Outcome-Conditioned Generation.} Unlike language modeling, where the output sequence receives a single reward value~\cite{PPO22}, users interact with recommended items through diverse engagement types—saving, sharing, commenting, or clicking. To effectively optimize across these varied engagement metrics, generative retrieval methods must account for this behavioral diversity~\cite{MultiBehaviorSASRec24}. Furthermore, practical deployment often requires specifying the target surface or context for the next recommendation, even when the user's historical sequence spans multiple surfaces.

To address these requirements, we design our model to accept conditioning signals at each generation step, enabling dynamic control over model behavior during inference. Specifically, we condition the output head $O$ on a desired set of outcomes, allowing the model to generate targeted recommendations. We implement this through learnable embeddings for each possible outcome type (e.g., engagement actions, surface constraints). By conditioning the generation process on these outcome representations, our model produces outcome-specific, personalized item representations:

\begin{equation}
    \hat{\mathbf{i}}_{u, t} = O(\mathbf{h}_{u,t}, \mathbf{c}_1, \mathbf{c}_2, \ldots, \mathbf{c}_k)
\end{equation}
where $\mathbf{h}_{u,t}$ is the hidden state at timestep $t$, $\mathbf{c}_1, \mathbf{c}_2, \ldots, \mathbf{c}_k$ are the conditioning embeddings for different outcome types (e.g., $\mathbf{c}_1 = \mathbf{e}_{\text{action}}$, $\mathbf{c}_2 = \mathbf{e}_{\text{surface}}$), and $\hat{\mathbf{i}}_{u, t}$ is the resulting conditioned item representation.

\subsection{Input Sequence Representation}
\label{sec:replearn}

Given the heterogeneous nature of user interactions $(i_j, a_j, s_j, t_j)$ in our interaction history $H(u, t_{\text{max}})$, we need to transform each element into appropriate representations for the transformer architecture. There are two main ways to represent items in a sequence: (1) by using dense embedding vectors~\cite{HSTU24, KangM18}, and (2) by learning a sequence of discrete IDs for each item, known as Semantic IDs~\cite{Tiger23, SemanticIDGeneral24}. We explored both approaches and ultimately chose to use dense embedding vectors. In our experiments, we observed that Semantic IDs frequently suffered from representational collapse—many items were assigned the same Semantic ID, as also noted in \cite{MultiBehaviorSASRec24, Hourglass24}—with tens of thousands of items colliding to the same ID at Pinterest's scale, and consistently produced lower evaluation metrics (see TIGER in Table~\ref{table:model_comparison})~\cite{UnifyingGenRetrieval24}. Additionally, since \pinrec generates continuous embeddings matched to real items via ANN search, it cannot hallucinate nonexistent items, a known failure mode of token-ID approaches. While optimizing performance with Semantic IDs remains an interesting direction for future work, our core contributions are representation-agnostic.

We now detail how each component of the interaction tuple $(i_j, a_j, s_j, t_j)$ is processed and integrated into our model architecture.

\textbf{Item Embeddings.} We generate dense embeddings for different entity types ($\tau \in \mathcal{T}$) using type-specific embedders that are learned end-to-end with the model. Each item type $\tau$ has a dedicated embedder $f_{\tau}$ that maps type-specific features to the shared representation space. While we describe embedders for Pins and search queries in this work, the approach can be extended to other entity types such as boards, users, or topics.

\textit{Pin Embeddings.} For Pins, we use the pre-trained OmniSage embedding~\citep{badrinath2025omnisage}, which captures visual and textual content along with the engagement via the Pin-board graph. The OmniSage embedding is then passed through 2-3 fully connected layers with ReLU activation and layer normalization, followed by $L_2$ normalization to produce the final Pin embedding $\mathbf{i}_{\text{pin}}$.

\textit{Query Embeddings.} For search queries, we leverage the pre-trained OmniSearchSage query representation~\citep{agarwal2024omnisearchsage} which captures semantic query understanding. This representation is passed through a similar MLP architecture (2-3 fully connected layers with ReLU activation and layer normalization) followed by $L_2$ normalization to generate the final query embedding $\mathbf{i}_{\text{query}}$.

\textbf{Temporal Embeddings.} For each interaction timestep $t_j$, we extract absolute timestamp and time since previous action, capturing both global temporal context and local session dynamics. Our encoder applies sinusoidal transformations at multiple frequencies. For absolute timestamps, we use predefined periods $P = \{p_1, p_2, \ldots, p_k\}$ to capture daily, weekly, and seasonal patterns: $\sin(t_{\text{abs}} / p_i), \cos(t_{\text{abs}} / p_i)$. For relative time differences, we use log-scale frequencies augmented with learnable phase parameters. 

Our item representations rely on embeddings and signals that are available for new items unseen during training. This enables \pinrec to generalize to unseen items---such as Ads Pins and Search queries---as we show in Sec.~\ref{sec:zero_shot}, highlighting the strong generalization of generative retrieval models.

\textbf{Sequence Construction.} The final input sequence to the transformer is constructed by concatenating item and temporal embeddings, then adding surface embeddings for each interaction. This creates a rich contextual representation that captures what users interacted with (item), when they interacted (temporal), and where the interaction occurred (surface) within the platform.

\subsection{Training Objective}
As described in Sec.~\ref{sec:intro}, \pinrec is trained in two stages: pretraining and fine-tuning. Below, we formulate the training objective for each stage. Both stages use the same softmax-based loss; they differ only in the target outputs used for supervision and negative examples.

During both training stage (pretraining and fine-tuning), we employ teacher forcing with a next-item prediction objective. The hidden states $\mathbf{h}_{u,t}$ are processed through the conditioned output head $O$ to generate item representations that predict future items in the sequence. For outcome-conditioned generation, we condition the output head on the actual values of the target item during training.  

Following \citet{Pinnerformer22}, we compute similarity between the predicted item representation $\hat{\mathbf{i}}_{u,t}$ and candidate item $\mathbf{i}_c$ as $s(\hat{\mathbf{i}}_{u,t}, \mathbf{i}_c) = \lambda \cdot \hat{\mathbf{i}}_{u,t}^\top\mathbf{i}_c - Q(\mathbf{i}_c)$, where $Q(\mathbf{i}_c)$ is a bias correction term estimated using count-min sketches \citep{cormode2005improved} to account for the non-uniform sampling of in-batch negatives.

The sampled softmax loss maximizes similarity to target items relative to negative samples $\mathbf{N}$:

\begin{align}
L_s(\hat{\mathbf{i}}_{u,t}, \mathbf{i}_{\text{target}}) = -\log \frac{\exp(s(\hat{\mathbf{i}}_{u,t}, \mathbf{i}_{\text{target}}))}{\exp(s(\hat{\mathbf{i}}_{u,t}, \mathbf{i}_{\text{target}})) + \sum_{\mathbf{i}_n \in \mathbf{N}} \exp(s(\hat{\mathbf{i}}_{u,t}, \mathbf{i}_n))}
\label{eqn:softmax_loss}
\end{align}

\textbf{Pretraining.}
We pretrain the model with next-token prediction on the vast amount of life long user history sequence $H(u, t_{\text{max}})$ (Sec.~\ref{sec:setup}), which contains actions from across all surfaces in Pinterest and excludes impression-only items (Fig.~\ref{fig:training}). We compute the loss in Eq.~\ref{eqn:softmax_loss} at each position in the user history sequence and average it across positions.

\begin{equation}
L_\theta^{\text{pre}}(H(u, t_{\max})) = \frac{1}{|H(u, t_{\max})|}\sum_{t=1}^{t_{\max}-1} L_s(\hat{\mathbf{i}}_{u,t}, \mathbf{i}_{u,t+1})
\label{eqn:pretrain_sequence_loss}
\end{equation}

\textbf{Fine-tuning.}
For surface-specific fine-tuning, we collect the feed session sequence \(F(u,s,t)\) for user \(u\) on surface \(s\) at time \(t\), together with the user’s history up to time \(t\) to avoid leakage, \(H(u,t)\) (Sec.~\ref{sec:setup}). To perform next-action prediction over both contexts, we construct a combined activity sequence \(G\) by concatenating \(H(u,t)\) with the \emph{action-only} entries from \(F(u,s,t)\):
\[
G(u,s,t)
=
H(u,t)\ \Vert\
\Big\langle (i_j, a_j, s, t_j)\ :\ (i_j,a_j,t_j)\in F(u,s,t)\ \wedge\ a_j\in\mathcal{A}\Big\rangle .
\]
We then compute the sequence loss on \(G(u,s,t)\) in the same way as in pretraining, averaging the per-position losses.

\begin{equation}
L_\theta^{\text{ft}}(G(u, s, t)) = \frac{1}{|G(u,s, t)|}\sum_{t=1}^{|G|-1} L_s(\hat{\mathbf{i}}_{u,t}, \mathbf{i}_{u,t+1})
\label{eqn:ft_sequence_loss}
\end{equation}

Additionally we add impression-only items in \(F(u,s,t)\) to the in-batch negative set $\mathbf{N}$ to serve as hard negatives (Fig.~\ref{fig:training}). We show in Sec.~\ref{sec:results} that these impression-based negatives improve performance.

\subsection{Inference}
\label{sec:inference}

Our \pinrec\ model serves as a candidate generator within the broader recommendation stack, producing diverse item embeddings. The model operates in an autoregressive fashion to generate multiple candidate embeddings, which are then used to retrieve actual items through approximate nearest neighbor search.

To generate an ordered set of item recommendations, we employ autoregressive generation on the transformer. The specific generation process varies based on the model configuration (Fig~\ref{fig:model}):

\textbf{Unconditional Generation.} Without outcome conditioning, generation is straightforward: at each step, the model produces a single item embedding, appended to the sequence along with the target surface embedding and a predefined temporal offset $o = 10s$. This updated sequence is then used to generate the next embedding.

\textbf{Outcome-Conditioned Generation.}
We generate one candidate embedding per outcome at each step, yielding multiple candidates per generation step. At each step, a single forward pass produces embeddings for all outcomes, similar to multi-task prediction heads. The key difference from standard multi-task learning is the autoregressive loop: we sample one outcome's embedding and feed it back as input to the next step (Fig.~\ref{fig:model}), so generation adapts across steps.

\textbf{Budget Allocation.} To retrieve $N$ total items, we first allocate retrieval budget across all generated embeddings. For unconditional models, the budget is distributed equally among all embeddings. For outcome-conditioned models, allocation is performed independently for each action type according to the specified budget constraints $\mathcal{B}$, ensuring that each condition receives its designated proportion of the total retrieval budget.

\textbf{Embedding Compression.} Given that autoregressive generation produces multiple output representations, we observe that many embeddings are often similar to one another. During ANN retrieval, these similar embeddings would collapse into overlapping sets of retrieved items, reducing diversity. To address this, we implement an embedding compression strategy: each generated embedding is merged into a previously uncompressed embedding (in generation order) if their cosine similarity exceeds a predefined threshold. When embeddings are compressed together, their allocated budgets are summed to determine the total retrieval budget for the compressed group.

This candidate generation process enables \pinrec\ to serve as a flexible first-stage retrieval system within Pinterest's recommendation infrastructure. The generated candidates can be further refined by downstream ranking models, allowing the system to benefit from both the generative model's ability to capture complex user patterns and the precision of specialized ranking algorithms. The outcome-conditioning makes the system well suited to the multi-objective optimization needed to customize recommendations for each surface.

\vspace{-3mm}
\subsection{Serving}

\begin{figure}[!t] 
    \centering
    \includegraphics[width=1.0\linewidth, trim=0 40 0 0, clip]{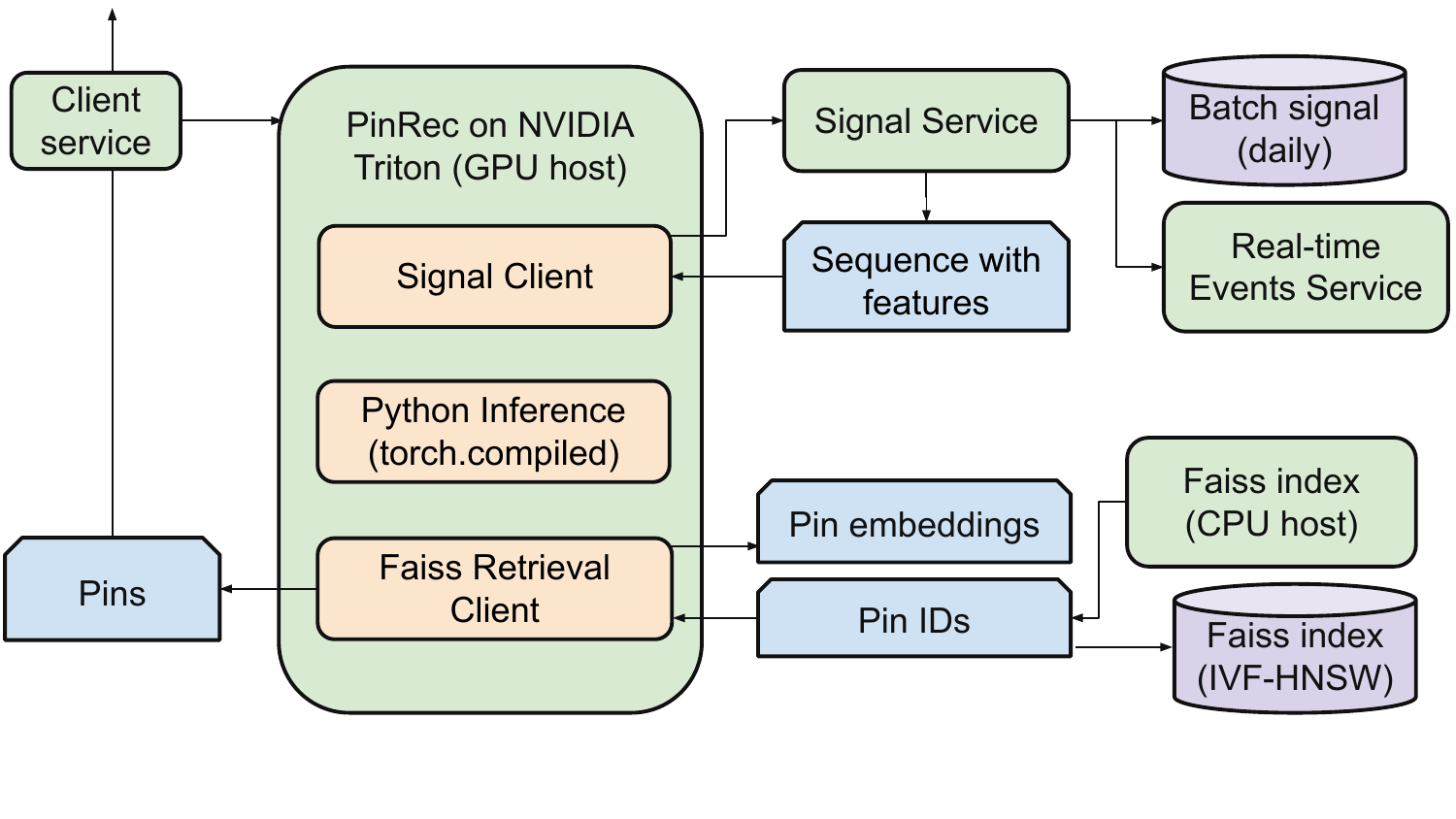}
    \caption{Serving flow for the \pinrec\ system. Green boxes represent services, blue rectangles represent transmitted data, purple cylinders represent indexed data, and beige boxes are steps within the \pinrec\ NVIDIA Triton ensemble.}
    \label{fig:serving}
    \vspace{-1mm}
\end{figure}

\pinrec\ uses an in-house optimized transformer implementation and served with NVIDIA Triton Inference Server's ensemble setting and Python backend to perform autoregressive generative retrieval. This multi-stage serving pipeline includes separate stages for fetching signals, featurizing Pins with Pin features, autoregressive inference for item generation, and Faiss retrieval (\autoref{fig:serving}).

\subsubsection{Signal Fetching}
Our Signal Service uses a Lambda architecture that combines both batch and real-time processing. The batch signal is updated daily through a Spark pipeline that processes the past year's positive user engagements and search queries, storing the sequences and features for each user ID. The real-time component is backed by a RocksDB key-value store \citep{kvstore} that efficiently retrieves user engagements occurring after the batch signal's cutoff time. The batch and real-time signals are deduplicated and returned to \pinrec. On surfaces with context (e.g., Search, Related Pins), model inference also includes the context (e.g., search query, closeup Pin) issued by the user at request time as the final element in the sequence. To limit transport costs, all embeddings associated with sequence elements are quantized to INT8 precision and later dequantized to FP16 for inference.


\subsubsection{Autoregressive Inference}
We serve the main \pinrec\ model for autoregressive generation on NVIDIA L40S GPUs. To generate an ordered set of item recommendations efficiently, we employ several optimizations for latency and cost efficiency:

\paragraph{CUDA Graphs.} To limit overhead from kernel launches and include other in-built PyTorch optimizations, we use CUDA Graphs with \texttt{torch.compile} to maximally compile the main components of the model.

\paragraph{KV Cache.} During the prefill stage of the transformer, the $K$ and $V$ tensors for the user's known sequence are cached in a per-batch-request KV Cache. On subsequent decode steps, these are reused, and only new sequence elements are processed to produce $K$ and $V$ for the additional sequence dimension.



\subsubsection{Faiss Retrieval}



Our nearest-neighbor retrieval involves a Faiss IVF-HNSW index running on CPU hosts, with inner product as the similarity metric. As with the main model, this is also served with NVIDIA Triton with batching enabled. We run $k$NN retrieval to fetch the allocated number of Pins for each embedding, with an overfetch factor for later de-duplication.


\subsubsection{Serving Efficiency}
\label{sec:serving_efficiency}
On the memory front, \pinrec uses about 1.6GiB of GPU memory including KVCache and CUDA graphs, without any significant memory usage increase for outcome conditioning. Excluding the KVCache and CUDA graphs, we use roughly the same amount of memory ($\approx$250 MiB) relative to existing two-tower retrieval.

In terms of latency, at 80 QPS, \pinrec-OC's latency was p50 40ms/p90 65ms, and \pinrec-UC's was p50 23ms and p90 46ms. Though the latency is roughly 3-4x higher (p50) than a traditional two-tower solution, we note that parallelizing it with other calls results in <1\% added end-to-end increase in latency.

\section{Offline Evaluation}
\label{sec:results}

In this section, we dissect the performance of \pinrec on generative retrieval across a range of offline evaluations. For the results presented, \pinrec\ uses a 12-layer transformer model (12 attention heads, 768 hidden dimensions). Our experimental setting includes interactions on three major surfaces for the Pinterest platform, namely Home Feed, Related Pins, and Search. We demonstrate that outcome-conditioned generation captures user intent more accurately, and that surface-specific fine-tuning further improves metrics on each surface. Importantly, these two techniques are \emph{complementary}: neither makes the other redundant. Each provides additional gains, and their benefits accumulate when used together.

\subsection{Offline Evaluation Metrics}

We construct the evaluation set using user activities disjoint from the training set and evaluate whether models can produce representations that retrieve future item embeddings from a set of $1M$ random items (the "index"). We use recall@$k$~\cite{Pinnerformer22, Pinsage18, UnifyingGenRetrieval24}, a standard performance metric measuring the proportion of target items correctly retrieved within the top-$k$ ranked items.

We evaluate model performance using a sequence generalization of recall@$k$ that accounts for the multi-embedding nature of our autoregressive generation. Unlike standard recall, which compares individual predictions to individual targets, our unordered recall metric compares sets of autoregressive generations to sets of future targets for each user.

Given output item representations
$\hat{\mathbf{I}}_u = \{\hat{\mathbf{i}}_{u,1}, \hat{\mathbf{i}}_{u,2}, \ldots, \hat{\mathbf{i}}_{u,m}\}$ and target item embeddings $\mathbf{I}_u = \{\mathbf{i}_{u,1}, \mathbf{i}_{u,2}, \ldots, \mathbf{i}_{u,n}\}$ for user $u$, along with random negative embeddings $\mathbf{N}$, we calculate the proportion of target items for which at least one predicted embedding retrieves it within the top-$k$ results:
\begin{align*}
    \label{eqn:recall}
    \begin{split}
    \mathrm{ur}_k(\mathbf{\hat{I}}, \mathbf{I}, \mathbf{N}) \coloneq \frac{1}{|\mathbf{\mathbf{\hat I}}|} \times \sum_{i=1}^{|\mathbf{\hat I}|} \mathbf{1}\left\{\min_{j,z \in [\Delta]}\left|\{ \mathbf{n} \in \mathbf{N} \mid \mathbf{\hat I}_{ij} ^\top \mathbf{n} \geq \mathbf{\hat I}_{ij}^\top \mathbf{I}_{iz}) \} \right| \leq k \right\}
    \end{split}
\end{align*}

where $\text{rank}(\mathbf{i}, \hat{\mathbf{i}}, \mathbf{N})$ denotes the rank of target item $\mathbf{i}$ when items in $\mathbf{N} \cup {\mathbf{i}}$ are sorted by cosine similarity to prediction $\hat{\mathbf{i}}$. This metric captures whether any of the generated embeddings successfully retrieves each target item, reflecting the practical scenario where multiple predictions can contribute to successful recommendation.

We perform autoregressive generation without teacher forcing, ensuring the transformer is not conditioned on target items during prediction, making evaluation realistic for deployment scenarios.

\begin{table}[t]
    \centering
    \caption{Comparison of baselines and \pinrec \ variants (unordered recall @ 10) across major surfaces at Pinterest.}
    \label{table:model_comparison}
    \begin{tabular}{l||ccc}
        \toprule
        Model & Home Feed & Related Pins & Search \\
        \midrule
        SASRec \citep{kang2018self} & 0.382 & 0.426 & 0.142 \\
        PinnerFormer \citep{Pinnerformer22} & 0.461 & 0.412 & 0.257 \\
        TIGER \citep{Tiger23} & 0.208 & 0.230 & 0.090 \\
        HSTU \citep{HSTU24} & 0.596 & 0.539 & 0.179 \\
        \pinrec-UC & 0.608 & 0.521 & 0.350 \\
        \pinrec-OC & \textbf{0.625} & \textbf{0.537} & \textbf{0.352} \\
        \bottomrule
    \end{tabular}
\end{table}

\subsection{Offline Results}

In this section, we compare \pinrec\ against existing models and quantify the impacts of outcome-conditioning, pretraining and fine-tuning.

\subsubsection{Model Evaluation} We compare \pinrec\ against existing sequence-based user modeling approaches, including Pinnerformer~\citep{Pinnerformer22}, SASRec~\citep{kang2018self}, and the leading generative recommendation methods TIGER~\citep{Tiger23} and HSTU~\citep{HSTU24}. These baselines span both paradigms, sequential user models (SASRec, PinnerFormer) and generative recommenders (TIGER, HSTU), representing the strongest published methods that scale to Pinterest's data volume. We evaluate different variants of \pinrec.
\begin{itemize}[align=parleft,left=0pt..1em]
    \item \pinrec-UC: We do not condition on anything (i.e., $\Phi = \emptyset$)
    \item \pinrec-OC: We condition on the target action (``outcome conditioned'') 
\end{itemize}

To make it easier and fair to evaluate many model architectures, we use the same training procedure for all methods; all models are trained from the activity sequence from all surfaces like \pinrec pretraining (Eq.~\ref{eqn:pretrain_sequence_loss}). We evaluate on user activity sequences from held-out users (i.e., users not seen during training). There is no fine-tuning process involved in this evaluation. Additional details for HSTU evaluation, the impact of negative sampling and model scaling effects are in Sec. \ref{appendix:HTSU_ablation}, \ref{appendix:negative_sampling} and \ref{appendix:scaling}, respectively.

Our comparison in \autoref{table:model_comparison} shows that \pinrec outperforms other baselines across the board. Importantly, outcome-conditioning demonstrates notable improvements over \pinrec-UC (unconditioned) in unordered recall @ 10 (e.g., \textbf{+2-3\%} on Home Feed and Related Pins).

\subsubsection{Fine-tuning Evaluation}
\label{sec:oc_fine_tuning}
We evaluate the impact of fine-tuning using surface-view logs. Since surface traffic and content evolve over time, we adopt a time split: we train on logs before a cutoff time and evaluate on logs collected after the cutoff, following standard industrial recommender evaluation protocols~\cite{PinFM25, Transact23}. This time-split evaluation set differs from the evaluation data used in Table~\ref{table:model_comparison}.

Table~\ref{table:training_comparison} presents a controlled ablation from the best configuration, \pinrec-OC \{FT, IN\}, removing one component at a time. We report relative change in recall, $\frac{\text{ablated} - \text{best}}{\text{best}}$. The best configuration uses outcome conditioning (OC), surface-specific fine-tuning with surface actions as additional positives (FT; Eq.~\ref{eqn:ft_sequence_loss}), and impression-only items as in-batch negatives (IN). We ablate as follows:
\begin{itemize}[align=parleft,left=0pt..1em]
    \item Remove OC: Remove outcome conditioning.
    \item Remove IN: Remove impression-only items from in-batch negatives.
    \item Remove IN, add IP: Replace impression negatives with impression positives (treat shown-but-not-engaged items as positives).
    \item Remove FT: Remove fine-tuning entirely (pretrained model only). Since IN is only available during fine-tuning, this also removes IN.
\end{itemize}

Outcome conditioning is the most critical component, with its removal causing -7.5\% to -12.6\% recall degradation across surfaces. Fine-tuning provides consistent gains of +2.0--4.5\%, supporting our claim that each surface benefits from adapting the model to its distinct user behavior patterns. Impression negatives yield a small additional improvement on Home Feed, while replacing them with impression positives hurts Related Pins and Search significantly, as it rewards exposure rather than intent.

\textbf{Together, these results show that outcome conditioning and fine-tuning are complementary: neither makes the other redundant, and their benefits accumulate when combined.}

\begin{table}[t]
    \centering
    \caption{Controlled ablation of \pinrec-OC \{FT, IN\} (recall @ 10). Each row modifies one component from the best configuration.}
    \label{table:training_comparison}
    \begin{tabular}{l||ccc}
        \toprule
        Ablation & Home Feed & Related Pins & Search \\
        \midrule
        \pinrec-OC \{FT, IN\} (best) & 0.0\% &  0.0\% & 0.0\% \\
        Remove OC & -12.6\% & -7.5\% & -11.9\% \\
        Remove IN & -0.4\% & +0.1\% & 0.0\% \\
        Remove IN, add IP & -0.4\% & -3.0\% & -6.8\% \\
        Remove FT & -4.5\% & -2.0\% & -2.5\% \\
        \bottomrule
    \end{tabular}
\end{table}

\begin{table}[t]
    \centering
    \caption{Percentage lift in recall@10 for PinRec-OC over PinRec-UC when matching outcomes and desired actions.}
    \label{table:oc_comparison}
    \begin{tabular}{l l c}
        \toprule
        Surface & Outcome & Recall Lift \\
        \midrule
        \multirow{2}{*}{Home Feed} & Save & +1.9\% \\
                               & Outbound Click                & +6.2\%  \\
        \midrule
        \multirow{2}{*}{Search} & Product Outbound Click & +16\% \\
                               & All other actions                & +8\%  \\
        \bottomrule
    \end{tabular}
\end{table}

\subsubsection{Outcome-Conditioning's Impact on Metrics} We assess whether conditioning on a desired action (e.g., “save”) leads to outputs that elicit that action. If our recommendations successfully control outcomes, we expect higher offline recall for the conditioned action. 

With \pinrec-OC\{FT, IN\} models, we allocate the entire budget to a single action and run recall only for the action (\autoref{table:oc_comparison}). We see larger improvements for rarer outcomes. For example, product outbound click is a rare but important search action, since Pinterest wants users to click through to products they intend to shop. In this case, \pinrec-OC delivers a 16\% lift.

\subsubsection{Autoregressive Generation Ablation}
We ablate the number of autoregressive (AR) generation steps to quantify the benefit of multi-step generation over a single-pass baseline. Table~\ref{table:ar_ablation} reports recall lift relative to 1-step generation, which produces multiple embeddings in a single forward pass without autoregressive feedback.

Autoregressive generation yields substantial gains across all surfaces, with Related Pins benefiting most (+71.4\% at 16 steps). This confirms that the AR loop, not just the multi-embedding output, drives PinRec's retrieval improvements, as the model adapts its generation to evolving user intent across steps.

\begin{table}[t]
    \centering
    \caption{Recall@10 lift from autoregressive generation steps, relative to 1-step (no AR feedback).}
    \label{table:ar_ablation}
    \begin{tabular}{l||ccc}
        \toprule
        AR Steps & Home Feed & Related Pins & Search \\
        \midrule
        1 (baseline) & 0.0\% & 0.0\% & 0.0\% \\
        4 & +5.9\% & +42.9\% & +8.8\% \\
        8 & +8.1\% & +57.1\% & +10.3\% \\
        16 & +8.8\% & +71.4\% & +11.1\% \\
        \bottomrule
    \end{tabular}
\end{table}

\section{Online A/B Experiments}

We perform online A/B experiments to evaluate \pinrec \ as a Candidate Generator (CG), measuring its impact on user engagement and its ability to tune the trade-off between business metrics. The items recommended by \pinrec \ are passed through (unmodified) ranking and blending stages on each surface, alongside other Candidate Generators (CGs), to produce the final results shown to users. We tested on the three major surfaces: Home Feed, Search and Related Pins.

\subsection{Trading Off Objectives: A Pareto Analysis} 
\begin{figure}[htb]
    \centering
    \begin{subfigure}{0.49\linewidth}
        \centering
        \includegraphics[width=\linewidth]{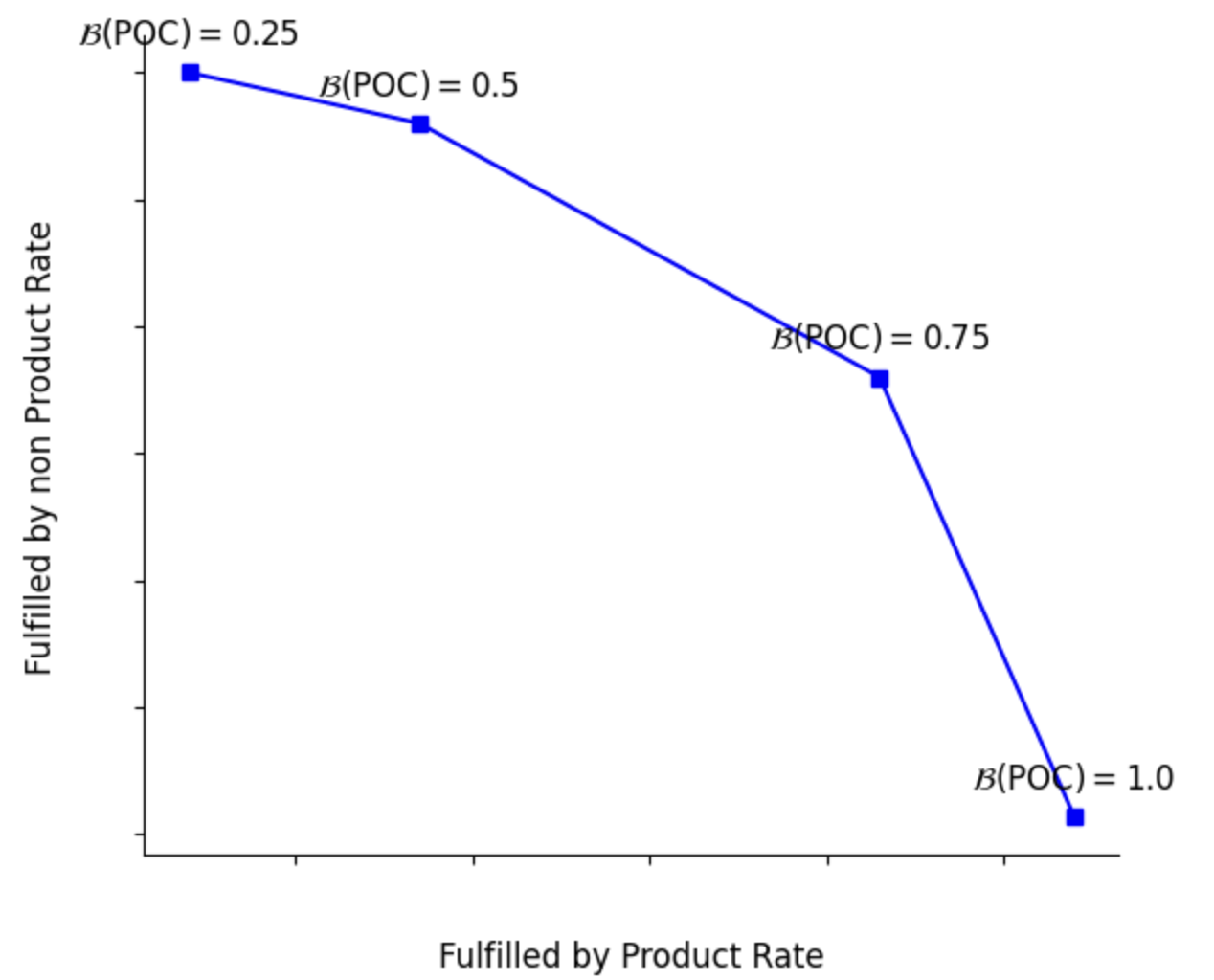}
        \caption{\pinrec-OC Search Goals}
        \label{fig:search_shopping_pareto}
    \end{subfigure}
    \begin{subfigure}{0.49\linewidth}
        \centering
        \includegraphics[width=\linewidth]{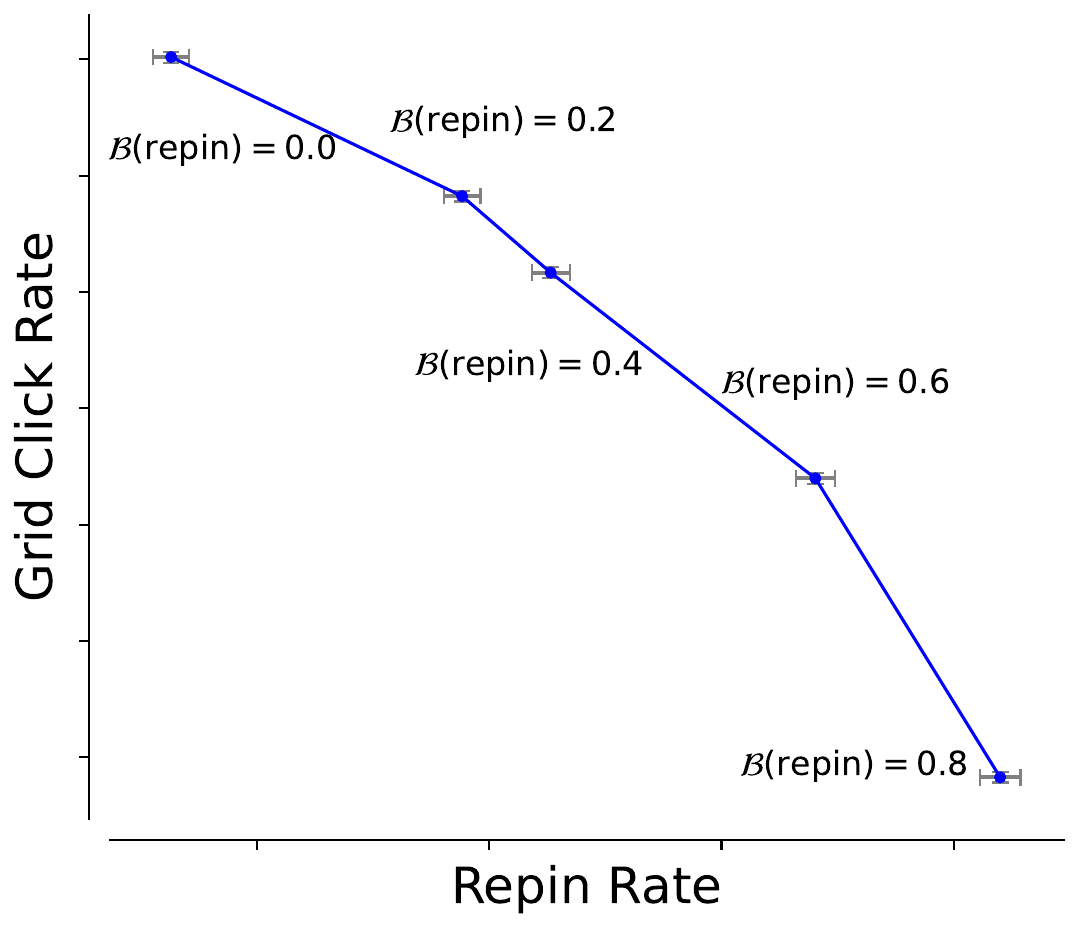}
        \caption{\pinrec-OC HF Goals}
        \label{fig:nonmt_pareto_cg}
    \end{subfigure}%
    \caption{Visualization of Pareto fronts for online engagement rates by outcome in surface-specific \pinrec \ variants.}
    \vspace{-0mm}
    \label{fig:combined_pareto_cg}
\end{figure}

To validate that \pinrec can efficiently control trade-offs between business objectives, we run five \pinrec-OC configurations with varying action budgets $\mathcal{B}$. We allocate a budget of $\mathcal{B}$ to outcome A and $1-\mathcal{B}$ to the other outcome.
\autoref{fig:combined_pareto_cg} shows the resulting Pareto frontier for the two outcome rates (normalized by impressions) produced by the \pinrec\ candidate generator alone; absolute values are omitted due to non-public information. These results show that \pinrec enables the business to balance multiple metrics by tuning the trade-off between objectives. Additional results are shown in \autoref{fig:combined_pareto_helium}.

\subsection{Surface A/B Tests}

\subsubsection{Metrics}
We use engagement actions such as saves, grid clicks, and time spent. Additionally, we count “fulfilled sessions / fulfillment,” where the user takes at least one action in the session, and “unfulfilled sessions,” where the user does not take any action. Note that \textbf{all reported metric in bold letters are statistically significant} at the 5\% level (two-sided test; $p<0.05$).

\subsubsection{Home Feed}

For Home Feed, we evaluated both \pinrec-UC and \pinrec-OC (\autoref{tab:online_hf}). While \pinrec-UC increases grid clicks, \pinrec-OC delivers larger gains in downstream actions and in fulfilled sessions, demonstrating the benefit of outcome-conditioning.

\begin{table}[t]
    \centering
    \caption{A/B experiments for \pinrec \ variants as candidate generators in Home Feed.}
    \label{tab:ab_results1}
    \begin{tabular}{llcc}
        \toprule
        Metric Type & Metric &  UC & OC \\
     \midrule
       \textit{Overall Metrics} &  Fulfilled Sessions  & +0.02\% & \textbf{+0.21\%} \\
        \midrule
        \textit{Action} & Site-wide Grid Clicks & \textbf{+0.58\%} & \textbf{+1.76\%} \\
        \textit{Metrics} & Home Feed Grid Clicks & \textbf{+1.87\%} & \textbf{+4.01\%} \\
        \bottomrule
    \end{tabular}

    \label{tab:online_hf}
\end{table}

\subsubsection{Related Pins}
\autoref{tab:related_ab_results} shows the metric lift from the \pinrec-OC on the Related Pins surface, with session-wide gains in fulfilled sessions and sitewide engagement. We also see cross-surface impact in Search: fulfillment increases by \textbf{+0.42\%} and saves by \textbf{+0.39\%}. \pinrec did not change Search retrieval or ranking; these gains are driven indirectly by better Related Pins recommendations leading to more search actions, highlighting \pinrec’s cross-surface learning.

\begin{table}[t]
    \centering
    \caption{A/B experiments for \pinrec \ variants as candidate generators in Related Pins.}
    \vspace{-1mm}
    \begin{tabular}{llc}
        \toprule
        Metric Type & Metric &  PinRec-OC \\
     \midrule
       \textit{Overall} &  Fulfilled Sessions  &\textbf{+0.21\%}  \\
        \textit{Metrics} & Time Spent & \textbf{+0.38\% } \\
        \midrule
        \textit{Action} & Site-wide Grid Clicks & \textbf{+0.76\%}  \\
        \textit{Metrics} & Sitewide Saves & \textbf{+0.26\%} \\
        \bottomrule
    \end{tabular}
    \vspace{-1mm}
    \label{tab:related_ab_results}
\end{table}

\subsubsection{Search}
\autoref{tab:search_ab_results} shows the results on Search A/B test. \pinrec-OC improves search metrics significantly relative to a control group, with increased session-wide search fulfillment (sessions resulting in positive interactions like saves or long clicks).

\begin{table}[t]
    \centering
    \caption{A/B experiments for \pinrec \ as candidate generator in Search retrieval.}
    \vspace{-0mm}
    \label{tab:search_ab_results}
    \begin{tabular}{llc}
        \toprule
        Metric Type & Metric & \pinrec-OC \\
        \midrule
        \textit{Overall Metrics} & Search Fulfillment Rate & \textbf{+2.24\%} \\
        \midrule
        \textit{Action Metrics}&  Search Saves & \textbf{+3.88\%} \\
        & Search Shares & \textbf{+5.30\%} \\
        \bottomrule
    \end{tabular}
\end{table}


\subsection{Zero-shot Retrieval}
\label{sec:zero_shot}
\pinrec represents items and queries using shared, broadly available feature backbones (e.g., OmniSage~\cite{badrinath2025omnisage} and OmniSearchSage~\cite{agarwal2024omnisearchsage}). Because these features are defined for a wide range of Pinterest entities, \pinrec can be applied to items or queries even when they are absent from the \pinrec training corpus.
We therefore evaluate \textit{zero-shot} retrieval, asking whether \pinrec can retrieve entities that never appear during \pinrec training.

Zero-shot capability is important in practice because Pinterest inventory and user intent evolve continuously: new Pins and ads are created every day, and new or tail queries emerge with limited interaction history. A retriever that generalizes zero-shot can serve such cold-start entities immediately, reducing reliance on frequent retraining and improving coverage for long-tail content.

After running \pinrec inference for the \pinrec\ candidate generator (CG), we take the model’s output embedding and perform nearest-neighbor search over other entity corpora:
\begin{itemize}
    \item \textbf{Related Searches:} retrieve search queries whose embeddings are closest to the user’s \pinrec embedding.
    \item \textbf{Ads:} retrieve ad Pins using the same nearest-neighbor retrieval approach.
\end{itemize}
For both the query and ads corpora, we precompute and index embeddings by running \pinrec inference with their respective features described in Sec.~\ref{sec:replearn}.

Table~\ref{tab:zero_ab_results} reports the results. Despite not being trained to predict engagement for these entities, \pinrec retrieves highly relevant candidates (e.g., lower cost per ad action). Moreover, this retrieval is substantially cheaper in production because \pinrec user-side inference is shared with organic CG. These results highlight the strong generalization of generative retrieval and support the case for a unified GR model.

\begin{table}[t]
    \centering
    \caption{A/B experiments for \pinrec \ as a zero-shot candidate generator.}
    \vspace{-1mm}
    \begin{tabular}{p{0.33\linewidth} p{0.33\linewidth} l c}
        \toprule
        Type & Metric &  PinRec-OC \\
     \midrule
       \textit{Related Searches} & Fulfilled Sessions  &\textbf{+1.66\%}  \\
        \textit{(Higher the better)} & Module Click Rate & \textbf{+4.04\% } \\
        \midrule
        \textit{Ads} & Cost Per Acquisition & \textbf{-1.83\%}  \\
        \textit{(Lower the better)} &  Cost Per Click & \textbf{-1.41\%} \\
        \bottomrule
    \end{tabular}
    \vspace{-1mm}
    \label{tab:zero_ab_results}
\end{table}

\section{Conclusion}
We introduced \pinrec, a unified generative retrieval system developed for Pinterest’s Homefeed, Search and Related Pins feed. To the best of our knowledge, this paper represents the first comprehensive study on deploying unified generative retrieval in a web-scale application. The incorporation of two key innovations --- outcome-conditioned generation and surface-specific fine-tuning --- enabled \pinrec to enhance several business metrics across multiple applications.

Several promising future directions can be pursued. First, applying generative approaches to ranking, as explored by Zhai et al.~\cite{HSTU24}, could be beneficial. Second, optimizing input sequences to the model, akin to prompt engineering in large language models, is another avenue worth exploring. Third, developing language model alignment techniques similar to ~\citet{Plum25} could enhance the use of engagement signals from users. Overall, this paper highlights a promising avenue for developing unified generative retrieval for industrial recommender systems.


\begin{acks}
We thank Xinyi Li, Hongtao Lin, Hanyu Li, and Aditya Mantha for their critical contributions. We are also grateful to our collaborators Divyansh Agarwal, Akanksha Baid, Bee-Chung Chen, Hung-Ju Chen, Bowen Deng, Stephanie Dewet, Huizhong Duan, Pong Eksombatchai, Dinesh Govindaraj, Kurchi Subhra Hazra, Henry Hu, Bella Huang, Se Won Jang, Saurabh Vishwas Joshi, Krishna Kamath, Sujay Khandagale, James Li, Cristian Lopez, Hanlin Lu, Lei Pan, Arun Sethupat, Dylan Wang, Ruijia Wang, Sai Xiao, Helen Xu, and Josie Zeng for their contributions and support.
\end{acks}



\bibliographystyle{ACM-Reference-Format}
\balance
\bibliography{references}

\appendix

\section{Appendix}

\subsection{Additional Analysis}

In this section, we perform additional experiments to quantitatively and qualitatively show that the design decisions made for \pinrec \ are justified and that it performs across the state-of-the-art architectures. Additionally, we provide more online analyses of the performance across different user cohorts and tail items.

\subsubsection{HSTU Architecture}\label{appendix:HTSU_ablation}

Though \pinrec \ relies primarily on the well-known transformer architecture, we note that our technique is completely architecture-agnostic. Specifically, outcome conditioning can be applied regardless of the specific architecture (e.g., transformer blocks) or algorithms used (e.g., attention), allowing them to be generic for any future use case of generative retrieval despite architectural or algorithmic improvements. To demonstrate this, we reproduce significant gains over baselines using the HSTU architecture \citep{HSTU24}, as opposed to the multi-head attention-based causal transformer. Note that we do not leverage the ranking task or any sequence sampling explored in \cite{HSTU24} as it is not relevant for our specific use case. For simplicity, we reproduce results on the outcome conditioned variant of \pinrec \ only, as we believe similar results can be extrapolated to other variants.

The results are shown in \autoref{table:comparison2}, which clearly shows that HSTU is roughly 3\% worse on Home Feed (effectively negating the improvement with respect to the unconditioned transformer-based \pinrec) and competitive on the related pins surface. However, there is a substantial drop in performance on the search surface, which is a critical component of the user experience at Pinterest. We believe that this may be related to specific architectural elements of HSTU.

\begin{table}[bth]
    \centering
    \caption{Comparison of HSTU and transformer architectures for \pinrec, where \pinrec-OC leverages the transformer as presented in the main text.}
    \label{table:comparison2}
    \begin{tabular}{l||ccc}
        \toprule
        Model & Home Feed & Related Pins & Search \\
        \midrule
        \pinrec-OC & 0.625 & 0.537 & 0.352 \\
        HSTU (OC) & 0.596 & 0.539 & 0.179 \\
        \bottomrule
    \end{tabular}
\end{table}

Though we observe a sharp drop in the search surface, HSTU performs roughly on the same order of magnitude as the traditional transformer architecture, demonstrating that our ideas perform across different architectures. This validates that our ideas are robust to architectures across the board. 

\subsubsection{Ablating Negative Sampling}\label{appendix:negative_sampling}

Following from \citet{Pinnerformer22}, we perform ablations on negative sampling, as it is a critical component of the chosen sampled softmax loss function. Though we ultimately end up with similar conclusions as \citet{Pinnerformer22}, we show these findings for maximum clarity regarding our design choices. 

\paragraph{Random Negatives} Though in-batch negatives are a critical component of our negative corpus, we construct a random negative corpus consisting of a random set of possible items. Given that our retrieval problem is ultimately across a random selection of pins, we wish for \pinrec \ to show retrieval capability across this corpus. We perform several ablations with random negatives, with in-batch negatives constant at a non-zero value. For these analyses, we focus on Home Feed (the most critical surface) and present results aggregated across all surfaces as well.

We show results in \autoref{table:comparison_rndneg2}, where there are moderate gains in unordered recall if there is a non-zero number of in-batch negatives present. Specifically, on the most critical surface of Home Feed, we see notable improvement over the baseline in terms of unordered recall.

\begin{table}[bth]
    \centering
    \caption{Percentage lift in unordered recall using 16K random negatives, relative to using only 8K random negatives.}
    \label{table:comparison_rndneg2}
    \begin{tabular}{l||c}
        \toprule
        Surface & \% Lift in Recall @ 10 \\
        \midrule
        Home Feed & +2.5\% \\
        All & +1.3\% \\
        \bottomrule
    \end{tabular}
\end{table}

To summarize, while random negatives are certainly an important piece of the puzzle, they cannot substitute in-batch negatives. When used in conjunction, scaling the number of negatives yields notable improvement, especially on Home Feed.

\paragraph{In-Batch Negatives} As shown in the previous subsection, disabling in-batch negatives entirely leads to significantly poorer metrics, showing their importance. Can we improve performance offline by scaling the batch size and consequently, the number of in-batch negatives? To examine this, we fix the number of random negatives and ablate the number of in-batch negatives through the batch size. 

\begin{table}[bth]
    \centering
    \caption{Percentage lift in unordered recall @ 10 through scaling the number of in-batch negatives (relative to baseline of 1x) with random negatives fixed.}
    \label{table:in_batch}
    \begin{tabular}{l||cc}
        \toprule
        \# In-Batch Negs. & \% Lift (HF) & \% Lift (all) \\
        \midrule
        1.5x & +4.2\% & +3.3\% \\
        2x & +8.4\% & +5.4\% \\
        12x & +18.3\% & +12.1\% \\
        \bottomrule
    \end{tabular}
\end{table}

We show results in \autoref{table:in_batch}, which demonstrates a clear trend that shows that a greater number of in-batch negatives and batch size leads to better performance, especially for the Home Feed surface. These are similar in directional improvements to random negatives, but we observe greater improvements with in-batch negatives, which are likely more relevant as negative examples.

\subsubsection{Ablating Parameter Scaling}\label{appendix:scaling}

One of the recent explorations within recommendation systems has been scaling to improve performance, in an attempt to replicate the success of these efforts for language modeling. To explore this hypothesis, we scale the number of parameters within \pinrec.

As presented in the main text, \pinrec \ (all variants) have roughly 100 million transformer parameters, corresponding to the base form or size of GPT-2. To ablate the effect of the size of the transformer, we attempt to follow the different scales employed by GPT-2 (base, medium, large, XL) as they have shown success in terms of the factors by which the embedding size, number of layers, and number of heads have scaled. Note that due to computational constraints, we reduce the scaling in terms of the number of layers, which increases GPU memory footprint significantly. Note that we present lift in offline unordered recall @ 10 for the outcome conditioned variant of \pinrec \ below, but we observe similar trends for the unconditioned model as well.

\begin{table}[bth]
    \centering
    \caption{Percentage lift in unordered recall through expanding the transformer at various GPT-2 scales relative to base (~100 million parameters).}
    \label{table:comparison4}
    \begin{tabular}{l||c}
        \toprule
        Model & \% Lift in Recall @ 10 \\
        \midrule
        Medium & -0.8\% \\
        Large & -0.9\% \\
        XL & +2.2\% \\
        \bottomrule
    \end{tabular}
\end{table}

Though we observe some gains in \autoref{table:comparison4} from scaling the number of transformer parameters to around 1 billion (XL-sized), these become intractable to serve in real time, requiring offline batch inference or significantly higher cost and more optimization. For the remaining sizes, there was no improvement observed actually. Hence, we opt to not use larger sized transformers.

\subsubsection{Tail User and Item Analysis}

One of the potential drawbacks of outcome conditioning would be for ``tail'' or ``fresh'' users, i.e., those which have lesser activity or less propensity for high-effort engagement. Critically, our method intends to optimize for outcomes across all cohorts of users, including such users. In the same vein, we wish to optimize for ``tail'' items, which may receive a lesser number of impressions, in an effort to distribute content to improve corpus coverage.

\begin{table}[ht]
\centering
\begin{subtable}[t]{0.45\textwidth}
\centering
\begin{tabular}{|l||cc|cc|} 
\hline
\multirow{2}{*}{User Cohort} & \multicolumn{2}{c|}{Grid Clicks} & \multicolumn{2}{c|}{Outbound CTs} \\
\cline{2-5}
 & \textbf{UC} & \textbf{OC} & \textbf{UC} & \textbf{OC} \\
\hline
New        & +2.60\% & +2.62\% & +3.98\% & +5.78\% \\
Resurrected & +1.56\% & +1.86\% & +1.98\% & +1.72\% \\
\hline
\end{tabular}
\caption{Online lift (\%) for grid clicks and outbound clickthroughers for new and resurrected users.} \label{tab:sub1}
\end{subtable}%
\hspace{1em}
\begin{subtable}[t]{0.45\textwidth}
\centering
\begin{tabular}{|l||c|c|}
\hline
Metric & \textbf{UC} & \textbf{OC} \\
\hline
Corpus Coverage (unique pins) & +6.2\% & +17.3\% \\
1-Imp Pin Rec. (tail items)   & +5.8\% &  +7.7\% \\
\hline
\end{tabular}
\caption{Relative lift (vs. 2-tower model) for tail item and corpus coverage metrics.}
\end{subtable}
\caption{Online experimental results for different user cohorts and tail item metrics, across \pinrec-UC and \pinrec-OC.} \label{tab:sub2}
\end{table}

In terms of new users, \pinrec-UC/OC lift grid clicks and outbound clickthroughers (outbound CTs) in \autoref{tab:sub1}. Critically, the lift for outcome conditioning-based modeling is either neutral or positive with respect to both control and the unconditioned variant of \pinrec, demonstrating that outcome conditioning does not suffer on such cohorts of users. In fact, across new users, OC performs better on outbound CTs.

To examine performance for tail items, we compare to a production 2-tower model in an online setting in \autoref{tab:sub2}. On corpus coverage, adding \pinrec-UC/OC surfaces 6.2\%/17.3\% more unique pins relative to the 2-tower. For long-tail items with 1-impression, \pinrec-UC/OC achieve 5.8\%/7.7\% absolute lifts respectively, relative to the aforementioned 2-tower model. Evidently, across the board, \pinrec \ surfaces more of the corpus and surfaces more tail items compared to control, with outcome conditioning showing even more significant lifts over control in both regards.

We believe the evidence suggests that our generative retrieval solution is equal to or better across inactive/new users and tail items relative to existing control, and perhaps more importantly, it shows that outcome conditioning can help in boosting performance in these criteria, which is one of its goals.


\begin{figure}[htb]
    \centering
    \begin{subfigure}{0.49\linewidth}
        \centering
        \includegraphics[width=\linewidth]{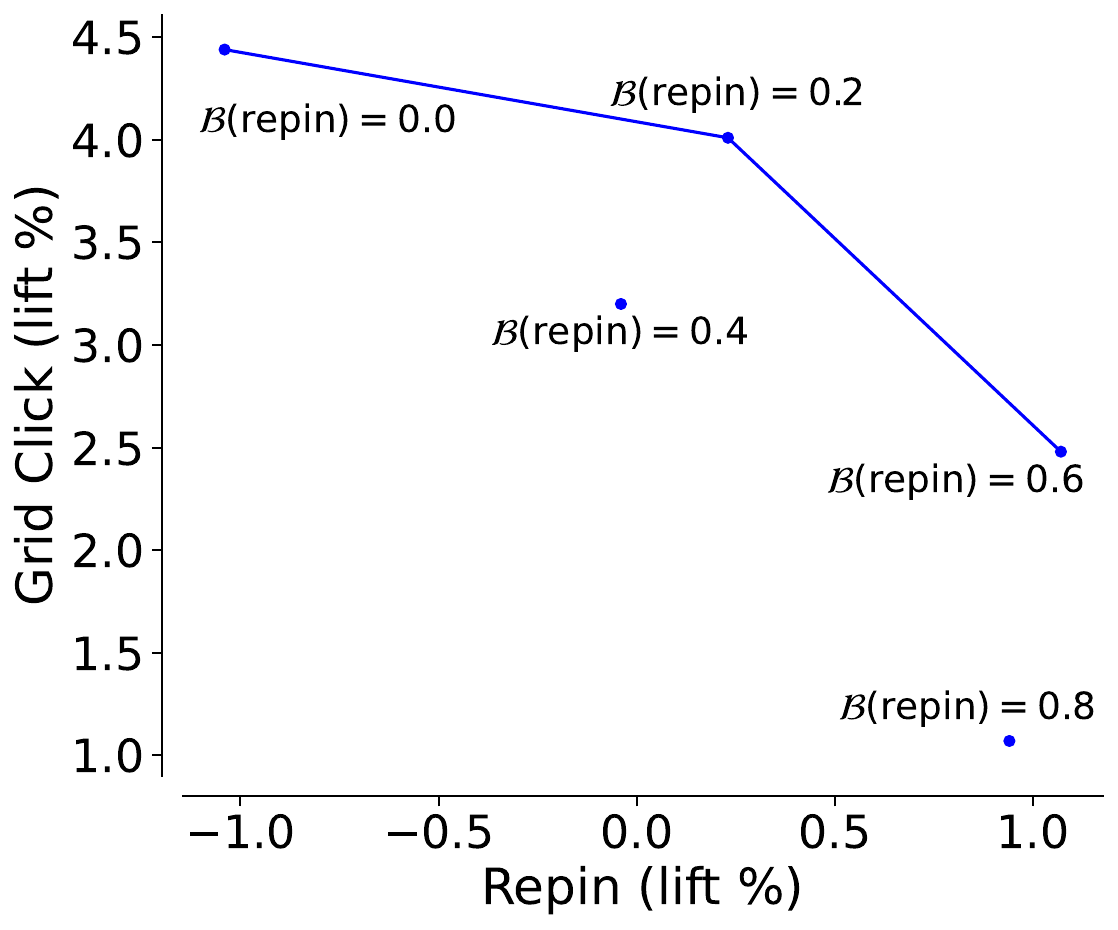}
        \caption{\pinrec-OC}
        \label{fig:nonmt_pareto_helium}
    \end{subfigure}%
    \vspace{-2mm}
    \caption{Visualization of Pareto fronts for online lift \% in Home Feed grid clicks versus repins for \pinrec }
    \label{fig:combined_pareto_helium}
\end{figure}

\subsubsection{Embedding similarity threshold tuning}
\label{sec:embedding_threshold}
For the embedding similarity threshold for embedding compression in Sec.~\ref{sec:inference}), we A/B tested values in the [0.75, 0.9] range for each surface. Our findings were: (1) results were qualitatively consistent across the different values, and (2) a threshold of 0.9 slightly outperformed other values for both Search and Related Pins, as reported in the paper. We will clarify this tuning procedure in the final version. We have not yet tuned this value for different user segments, but agree that this would be an interesting future research.

\subsubsection{Embedding Diversity Analysis}
\label{sec:embedding_diversity}
To quantify retrieval diversity, we measure the number of unique embeddings after compression (cosine similarity threshold of 0.85) across ablation variants. More unique embeddings query distinct regions of the ANN index, retrieving less overlapping candidate sets. Table~\ref{table:diversity_ablation} reports the relative change from the best configuration.

OC produces 2--12x more unique embeddings than the unconditioned variant. Fine-tuning increases diversity for exploratory surfaces (Home Feed, Related Pins) but decreases it for Search, where users have specific intent and narrower results are appropriate. This is corroborated by online corpus coverage results in Table~\ref{tab:sub2}.

\begin{table}[t]
    \centering
    \caption{Relative change in unique embeddings after compression, compared to \pinrec-OC \{FT, IN\}.}
    \label{table:diversity_ablation}
    \begin{tabular}{l||ccc}
        \toprule
        Ablation & Home Feed & Related Pins & Search \\
        \midrule
        \pinrec-OC \{FT, IN\} (best) & 0.0\% & 0.0\% & 0.0\% \\
        Remove OC & -67.3\% & -91.4\% & -92.9\% \\
        Remove FT & -37.8\% & -35.3\% & +13.1\% \\
        Remove IN & -31.8\% & +1.6\% & -1.0\% \\
        \bottomrule
    \end{tabular}
\end{table}

\subsubsection{Temporal offset value tuning}
\label{sec:temporal_offset}
\textbf{Inference:} We experimented with different values of $\delta$ for each action at inference time. For the results reported in this paper, we sampled repins and grid-clicks at $\delta \in \{0, 25, 50\}$ (i.e., 0, 25, and 50 discrete timesteps ahead). For all other outcomes, we sampled only the immediate next timestep ($\delta = 0$). We also tried sampling every 5 timesteps from 0 to 50, but this generated too many embeddings for ANN search.

\textbf{Training:} We randomly selected 4 training targets with $\delta$ sampled from the interval $[0, 25)$. For evaluation, we compared the model outputs at $\delta \in \{0, 10\}$ and measured recall@10 for predicting the ground-truth action under different Timestep Gaps (TG). TG = 0 corresponds to immediate-term recall, while TG = 30 reflects long-term recall. Compared to $\delta = 0$, $\delta = 10$ achieved 60\% higher long-term recall with only a 2\% decrease in short-term recall, indicating a better overall trade-off; therefore, we use $\delta = 10$.

\end{document}